# CBSE CASE ENVIRONMENT

*Luiz Fernando Capretz*
Department of Electrical & Computer Engineering
University of Western Ontario
1201 Western Road, London, N6G 1H1, CANADA
lcapretz@uwo.ca

**ABSTRACT**

With the need to produce ever larger and more complex software systems, the use of reusable components has become increasingly imperative. Of the many existing and proposed techniques for software development, it seems clear that components-based software engineering (CBSE) will be at the forefront of new approaches to the production of software systems, and holds the promise of substantially enhancing the software development and maintenance process. The required features of a CASE environment suitable for component reuse will be put forward in this paper.

**1 INTRODUCTION**

Matured engineering disciplines have handbooks that describe successful solutions to well-known problems, or a family of related problems, thus minor percentage in performance value gained by starting from scratch typically is not worth the cost [1]. A component is a self-contained piece of software that provides clear functionality, has open interfaces, and offers plug-and-play services. Component-based software engineering (CBSE) is expected to have a significant impact on the software industry, and hopefully on how software engineers construct systems, so this technique is here to stay [2]. There are initiatives in that direction, such as Microsoft COM+ [3], SUN JavaBeans [4], IBM Component-Broker [5] and CORBA [6], among others.

A CASE environment can provide computer-aided support for CBSE through a set of tools which form that environment, and can bring many improvements in software quality and efficiency for software production. Without such supporting tools (or strong management pressure!) there is no easy means of enforcing the use of the ideas behind a particular methodology. Software engineers may then merely pay lip-service to the adoption of the methodology, at best resulting in a minimal improvement in productivity, at worst putting the whole software development in jeopardy.

The high cost and complexity of software development and maintenance, and the growing need for reusable software components, are some of the factors stimulating research into better software methodologies and CASE environments. Of course, the aim of a methodology is to improve some (or all!) of the quality, reliability, feasibility, cost-effectiveness, maintainability, management and engineering of software, and these obvious desires will not be discussed further here. But actually putting a methodology into practice is almost impossible without a set of tools which gives automated support for that methodology. Thus, one requirement for CASE tools is that they support and promote a software development methodology by sustaining and enforcing the steps, rules, principles and guidelines dictated by the software development process.

**2 SEAMLESS SOFTWARE DEVELOPMENT**

The creation of software is characterized by change and instability, and therefore any diagrammatic representation of the seamless model should consider overlapping and iteration between its phases. However, a consensus may be drawn on the phases pertinent to a seamless life cycle. Although the main phases may overlap each other and iteration is also possible, the planned phases are: system analysis, domain analysis, design and implementation. Maintenance is an important operational phase, in which bugs are corrected and extra requirements met.

Figure 1 displays a pictorial representation of how the system analysis, domain analysis, design, implementation and maintenance phases proceed iteratively over time and how reuse of components from the reusable library is taken into consideration within that software life cycle model [7]. Reusability within this life cycle is smoother and more effective than within the traditional models because it integrates at its core the concern for reuse.

A feature of this software development model is the emphasis on reusability during software creation, and the production of reusable components meant to be useful in future projects. This is naturally supported by the object-oriented paradigm due to inheritance and encapsulation. Reusability also implies the use of composition techniques during software development. This is achieved by initially selecting reusable components and aggregating them, or by refining the software to a point where it is possible to pick out components from reusable software libraries.

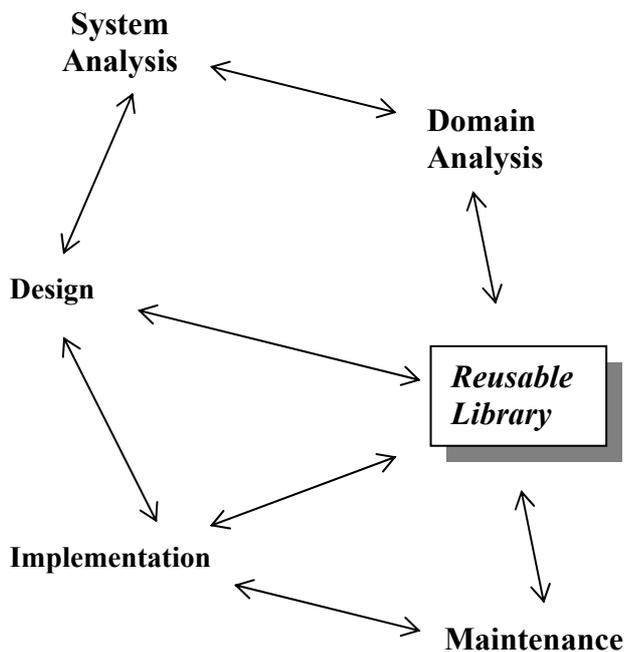

Figure 1: *Seamless Software Development*

## 3 A CASE FOR CBSE

The process of software system design has many facets and stages, spanning high-level design early in the process down to low-level design which is closely related to the implementation step of actually writing code. At different stages, the designer is likely to be manipulating different kinds of information, that is, a high-level design might concentrate on aspects of the classes and frameworks whereas lower-level design would need to consider interactions between objects which are instances of those classes and frameworks. Thus a set of notations will be required to permit the different sorts of information in a design to be captured and presented.

It is widely recognized that graphical notations have clear advantages over textually-based notations. Thus it seems paramount that a set of graphical notations is used to represent the different concepts within a methodology, with visually different notations being used to represent conceptually different ideas. For instance, the representation of a class should be different from that of an object. Such a set of graphical notations could be supported by one single, complex CASE tool, which tried to present one diagram encompassing all of the notations. However, with the complexity inherent in the design of a software system, it is most unlikely that a single diagram could be made to be comprehensible (or even displayable!) even if the CASE tool itself could be constructed. This therefore suggests that the requirement for a CASE environment is for a set of tools to support the design notations, each tool specialized to cope with one particular notation and each tool presenting just one aspect of a design, for example, one tool presenting the class hierarchy while another tool presenting a picture of object interactions.

Even though a set of tools is required, it is clear that the information manipulated by each tool is going to be interrelated. A set of tools isolated from each other will not be conducive to supporting the design process. Thus the requirement is that the tools are integrated together, to permit the designer to interchange information freely between one tool and another, and also to allow the designer to switch easily between tools as the needs arise.

The ability for a software engineer to navigate around is also vital as software reusability is something that a CASE environment must promote. A designer must be able to browse through already-captured parts of previous designs to try to see whether any components from prior work can be reused. However, one of the major problems which designers are faced with in trying to reuse software is the difficulty of finding reusable components, once such components have been produced. This is primarily because few mechanisms are available to help identify and relate components. In order to provide more convenient reuse, the question of which kinds of mechanisms might help solve this problem arises. The answer is typically couched in terms of finding components which provide specific functionality, from a library of potentially reusable components linked through relationships which express their semantics and functionality as a framework. Clearly, what is needed are techniques to create, classify and relate components, and CASE tools which help to store, select and retrieve potentially reusable components.

The CASE environment should also support consistency and completeness checking to ensure that the principles of the software process are being properly obeyed and that mistakes by the designer can be identified. Such checking could take place automatically within the tools. However, it is rare for the design stages to be completely separated from each other, with one level being completely finished before the next level down is commenced. More often than not, iterations around a design take place even though a part of the design is not complete - indeed, the ability to build up a design from an outline framework into which further details are filled in when other issues have been considered is a useful requirement.

The power of abstraction, which is at the heart of the design process, is to be able to ignore details until absolutely necessary. Thus it is more appropriate for checking and consistency tools to be separate from the other tools, permitting the designer to manipulate incomplete notations without continual warnings from automatic checks provided by tools. Checks can then be explicitly applied by the designer at appropriate points in the design process, for instance when the designer feels that some parts are complete.

The interface to the tool set is perhaps the most important factor in producing a CASE environment that is acceptable, especially when graphical notations are being manipulated. Even with user-friendly interfaces, it is also important that the tools have common interfaces to the extent that is possible. An environment with *n* tools with *m* different friendly interfaces is not going to be practicable. Tools which have common behaviour, such as the tools which are manipulating different graphical notations for the methodology, should adopt a common interface and support a consistent view of the information independent of which tool is being used at a particular moment.

All of the above requirements have been addressing aspects concerned with supporting the design methodology. Anothe rimportant aspect of software engineering concerns project management issues, and this is an area which also needs to be addressed within the CASE environment; the interested reader is referred to another work where such issues are discussed in depth [8].

## 4 SOME CBSE TOOLS

When designing a software system, a designer is faced with considering general and specific parts of a design. The general parts primarily encompass classes, since those are general components which could be added in (or reused from) a generic framework. The design of classes should therefore be with generality and reuse in mind, with perhaps less emphasis on satisfying the specific needs of the application that is being designed. In contrast, the specific parts of a design are those parts that turn a general set of components (e.g. classes) into a specific software design for a particular application, for instance by defining the instantiation and composition of objects from those classes and the algorithms which define the exact pattern of interactions between those objects within the generic framework. For example, one can have the design of a framework for dealing with text strings in a windowing environment. One specific application could use that framework to design a text-formatting program; another application could use the same framework to define a text editor.

Software engineers must recognize this distinction between general and specific parts of a design, and a design model should be regarded as being comprised of two parts:

1. An *information model* which represents the general aspects of a design.

2. A *behaviour model* which represents the application-specific parts of the design.

The information model is comprised of a global view of the static representation of components of the software system, i.e. classes and class hierarchies. The behaviour model is concerned with the dynamic relationships between objects, showing what objects are instantiated, how objects are composed and how they will interact in the specific application.

This distinction between the generic and specific aspects of a design is an important distinction which helps separate component-based software development from traditional paradigms . The idea of being able to classify parts of a design as generic, and hence potentially reusable, is a powerful reason for maintaining this distinction and indeed for spending more time on the general aspects of the design than might really be needed for a specific application. Thus, it is important to have tools which support the construction of separate information and behaviour models, yet maintaining the inter-tool relationships.

Furthermore, the nature of component-based software development by itself can affect the way that CASE environments are built because some tools are quite related to this paradigm, in that, they must manipulate large number of classes and objects as potentially reusable components. Therefore, some supportive tools are necessary, for instance:

- Library management tools to allow the searching for potentially reusable components such as classes and frameworks.

- Browsers to facilitate navigation through frameworks, storage and recovery of class hierarchies, and definition and visualization of classes in terms of their interface, attributes and operations.

- Checkers to examine design consistency and completeness. These tools should give some freedom to designers, letting them make temporary omissions and then checking for such omissions later in the design process.

The tools also handle the problem of feedback information (which is a vital part of a design) and help provide documentation for a software system via a report generator which basically produces reports from the descriptions of the design model.

When a seamless representation model is used to describe information, it has the added advantage of supporting traceability between software life cycle phases throughout software development because that model deals with uniform concepts (classes and objects) continuously, that is, the same concepts can be carried from the system analysis and design phases to the implementation and maintenance phases, even though the concepts change as they gain additional details during the later stages.